\documentclass{PoS}

\title{Realisation of a low frequency SKA Precursor: The Murchison Widefield Array}

\ShortTitle{The Murchison Widefield Array}

\author{\speaker{S.J. Tingay}\\
        International Centre for Radio Astronomy Research - Curtin University, Perth, Australia\\
        E-mail: \email{S.Tingay@curtin.edu.au}}

\author{R. Goeke, J.N. Hewitt, E. Morgan, R.A Remillard, C.L. Williams\\
        MIT Kavli Institute for Astrophysics and Space Research, Canbridge, USA}

\author{J.D. Bowman\\
        Arizona State University, Tempe, USA}
        
\author{D. Emrich, S.M. Ord, T. Booler, B. Crosse, D. Pallot, W. Arcus, T. Colegate, P.J. Hall, D. Herne, M.J. Lynch, F. Schlagenhaufer, S. Tremblay, R.B. Wayth, M. Waterson\\
        International Centre for Radio Astronomy Research - Curtin University, Perth, Australia}
        
\author{D.A. Mitchell, R.J. Sault, R.L. Webster, J.S.B. Wyithe\\
        The University of Melbourne, Melbourne, Australia}
        
\author{M.F. Morales, B.J. Hazelton\\
        University of Washington, Seattle, USA}
        
\author{A. Wicenec, A. Williams\\
        ICRAR - University of Western Australia, Perth, Australia}
        
\author{D. Barnes\\
        Swinburne University of Technology, Melbourne, Australia}
        
\author{G. Bernardi, L.J. Greenhill, J.C. Kasper\\
        Harvard-Smithsonian Center for Astrophysics, Cambridge, USA}
        
\author{F. Briggs, B. McKinley\\
        The Australian National University, Canberra, Australia}
        
\author{J.D. Bunton, L. deSouza, R. Koenig, J. Pathikulangara, J. Stevens\\
        CSIRO Astronomy and Space Science, Australia}
        
\author{R.J. Cappallo, B.E. Corey, B.B. Kincaid, E. Kratzenberg, C.J. Lonsdale, S.R. McWhirter, A.E.E. Rogers, J.E. Salah, A.R. Whitney\\
        MIT Haystack Observatory, Westford, USA} 
        
\author{A. Deshpande, T. Prabu, A. Roshi, N. Udaya-Shankar, K.S. Srivani, R. Subrahmanyan\\
        Raman Research Institute, Bangalore, India}     
        
\author{B.M. Gaensler\thanks{Gaensler, Briggs, McKinsley, Mitchell, Tingay, Tremblay, Wayth, Webster and Wyithe are affiliated with the ARC Centre of Excellence for All-sky Astrophysics (CAASTRO)}\\
        University of Sydney, Sydney, Australia}
        
\author{M. Johnston-Hollitt\\
        Victoria University of Wellington, New Zealand}
        
\author{D.L. Kaplan\\
        University of Wisconsin--Milwaukee, Milwaukee, USA}
        
\author{D. Oberoi\\
        National Centre for Radio Astrophysics, Pune, India}

\abstract{The Murchison Widefield Array is a low frequency (80 - 300 MHz) SKA Precursor, comprising 128 aperture array elements distributed over an area of 3 km diameter.  The MWA is located at the extraordinarily radio quiet Murchison Radioastronomy Observatory in the mid-west of Western Australia, the selected home for the Phase 1 and Phase 2 SKA low frequency arrays.  The MWA science goals include: 1) detection of fluctuations in the brightness temperature of the diffuse redshifted 21 cm line of neutral hydrogen from the epoch of reionisation; 2) studies of Galactic and extragalactic processes based on deep, confusion-limited surveys of the full sky visible to the array; 3)  time domain astrophysics through exploration of the variable radio sky; and 4) solar imaging and characterisation of the heliosphere and ionosphere via propagation effects on background radio source emission.  This paper will focus on a brief discussion of the as-built MWA system, highlighting several novel characteristics of the instrument, and a brief progress report (as of June 2012) on the final construction phase.  Practical completion of the MWA is expected in November 2012, with commissioning commencing from approximately August 2012 and operations commencing near mid 2013.  A brief description of recent science results from the MWA prototype instrument is given.}

\FullConference{Resolving the Sky - Radio Interferometry: Past, Present and Future - RTS2012\\
		April 17-20, 2012\\
		Manchester, UK}

\begin{document}

\section{Introduction}
The Murchison Widefield Array (MWA) is the only Square Kilometre Array (SKA: \cite{ska}) Precursor telescope at low radio frequencies.  An SKA Precursor is a recognised SKA technology demonstrator located at one of the two sites that will host the SKA, the Murchison Radio-astronomy Observatory (MRO) in the Murchison region of Western Australia and the Karoo region of South Africa's Northern Cape.  The MWA is sited at the MRO, along with a second SKA Precursor, the Australian SKA Pathfinder (ASKAP: \cite{jon08}; \cite{jon07}).  The MeerKAT SKA Precursor will be located at the South African site\footnote{http://public.ska.ac.za/meerkat}.  The MWA is sited at the MRO due to the extremely low levels of manmade radio frequency intereference in this area of Western Australia, particularly within the FM band (87 - 108 MHz) encompassed by the MWA at the low end of its operating frequency range.

The MWA and ASKAP are complementary, in that they operate in different frequency ranges (MWA: 80 - 300 MHz; ASKAP: 0.7 - 1.8 GHz) and employ different antenna technologies (MWA: aperture arrays (tiles); ASKAP: dishes plus Phased Array Feeds).  The combined MWA and ASKAP technology specifications almost fully sample the roadmap technologies for the SKA \cite{ska} at a single radio quiet location.  In particular, the MWA explores the so-called large-N and small-D array concept that will be utilised for the SKA, with large numbers of small receiving elements providing a large field of view on the sky and therefore high sensitivity over wide fields, equating to high survey speed, a key metric for SKA science \cite{skascience}.

While only three instruments have SKA Precursor status, a number of other instruments have SKA Pathfinder status (SKA technology demonstrator but not on a candidate SKA site) and are also feeding information into the SKA design and costing process.  The most notable of the SKA Pathfinders in the low frequency regime is LOFAR, built in The Netherlands \cite{lofar}.

All three of the MWA, ASKAP and MeerKAT are planning, or have operated, prototype instruments.  ASKAP is building the six antenna Boolardy Engineering Test Array (BETA\footnote{http://www.atnf.csiro.au/projects/askap}).  MeerKAT is being preceeded by KAT7\footnote{http://public.ska.ac.za/kat-7}, a seven antenna array.  For the MWA, a 32 tile test array operated for two years from 2009, allowing the assesment of a number of generations of prototype hardware as well as several trials of MWA system integration.  The MWA 32 tile array also undertook science-quality astronomical observations, in order to demonstrate on-sky performance.  A number of science results have been, or will soon be, reported from the prototype array data \cite{Williams-etal.2012, obe11}.  Operation of the 32 tile prototype array ceased in late 2011, in preparation for the start of construction for the final MWA instrument in early 2012.  A full description of the MWA science goals is given by \cite{bow12}.

Previously, \cite{Lonsdale-etal.2009} described the conceptual underpinnings of the MWA design, the advantages of a large-N, small-D architecture for radio interferometers and some of the challenges inherent in the data processing for this type of telescope.  \cite{Lonsdale-etal.2009} also provided brief descriptions of the initial plans for MWA hardware and data processing elements, as well as the science goals for the MWA.

Since the \cite{Lonsdale-etal.2009} paper was published, a number of design and construction considerations, that were uncertain at the time of publication, have been finalised.  In particular the originally envisaged 512 tiles has been re-scoped to 128 tiles.  This paper will give a brief summary of the as-built MWA system overview, a full description of which will appear elsewhere \cite{tin12}, a summary of results from the prototype array, and a brief report on current progress with the final construction of the MWA.  

Updates and news items for the MWA project are regularly posted at:\\ http://www.facebook.com/Murchison.Widefield.Array

\section{System overview of the as-built MWA}

The MWA signal path starts with a dual-polarisation dipole antenna, approximately a square metre of collecting area at $\sim$150 MHz.  Sixteen of these antennas are configured as an aperture array on a regular 4$\times$4 grid (with spacing 1.1 m); their signals are combined in an analog beamformer to provide coarse pointing capability, produces two wideband analog outputs representing orthogonal X and Y linear polarisations.  We refer to this configuration as an antenna tile and analog beamformer.

The four science drivers for the MWA lead to a desire for an aperture plane baseline distribution which has a dense core and a smooth $r^{-2}$ radial distribution.  The inner part of the array therefore has 50 antenna tiles uniformly distributed over a 100 metre diameter core, with a further 62 tiles distributed over a 1.5 km diameter circle (\cite{bea12}).   The final 16 tiles are placed on a 3 km diameter circle to optimise the highest angular resolution imaging.

A host of practical considerations lead us to combine the signals from eight tiles into a single receiver; we thus deploy 16 receivers distributed over our landscape such that no tile is more than 500 metres removed from its associated receiver.  A receiver filters the two analog signals from each tile to a bandpass of 80 to 300 MHz, Nyquist-samples the signals with an 8-bit A/D converter, and digitally filters the result into 24$\times$1.28 MHz frequency channels which form a selectable, usually but not necessarily contiguous, 30.72 MHz frequency band.  The receivers themselves are a mixture of high technology and low, combining high-speed, and hence high power usage, digital circuitry with mechanical cooling.  They are powered from a standard 240 V mains circuit, and send out their digital data streams on three 2.1 Gbps fibre optic links.  One of our early design decisions was to provide spare power and fibre connections to most receiver locations so that future instrument development could share our existing infrastructure at low incremental cost.  The excess infrastructure could support a doubling of the array to 256 tiles, or else support completely different instrumentation alongside the MWA facility.

About 5 km from the centre of the array lies a building provided by CSIRO, which we share with the ASKAP project.  It has power, EMI shielding, and water-cooled equipment racks for our signal processing hardware.  The data streams from all 16 receivers meet here and each of the coarse frequency channels is filtered by dedicated hardware into 128$\times$10 kHz fine channels.  A cross-multiply stage implemented in software using general purpose graphical processing units (GPGPUs) processes, averages in time and frequency space, and outputs its results in 768$\times$40 kHz channels over the 30.72 MHz bandwidth with 0.5 s resolution.  This 2.25 Gbps stream of correlation products is then processed by the Real Time System (RTS) to produce real-time calibrated images every 8 s.  The RTS also runs on the same GPGPU machine that performs the correlator cross-multiply.  We have enabled the ability for full visibility data to also be saved, to allow for the possibility of post-observation processing.  The RTS output and the $uv$ data will be transmitted on a dedicated 10 Gbps optical fibre connection to the Pawsey HPC Centre for SKA science in Perth, where 15 PB of data storage capacity has been reserved over a 5 year period, to accommodate the MWA data archive.

The MWA runs a monitor and control system to schedule observations and monitor system health and parameters of use for data processing.

Supporting the MWA instrument is the underlying infrastructure, including the reticulation of fibre and power around the array and the connection to MRO site-wide services provided by CSIRO.

\section{Results from the 32 tile prototype array}

The MWA project operated a 32 tile prototype instrument between 2009 and 2011.  The primary purpose of the prototype was to allow the iteration of design and implementation of the various MWA sub-systems.  In addition, where possible, science observations were supported with the prototype, which has resulted in a number of recent publications.  As it turned out, the prototype was far more scientifically capable than we could have hoped.  Selected results that are already publically available are briefly reviewed here.  A further five publications derived from prototype array data are currently in preparation.

\cite{obe11} presented the first spectroscopic images of solar radio transients from the prototype MWA, observed on 2010 March 27. The observations spanned the instantaneous frequency band 170.9 $-$ 201.6 MHz and, though the observing period was characterised as a period of ``low" to ``medium" activity, one broadband emission feature and numerous short-lived, narrowband, non-thermal emission features were evident in the data. The data represented a significant advance in low radio frequency solar imaging, enabling us to follow the spatial, spectral, and temporal evolution of events simultaneously and in unprecedented detail. The rich variety of features seen in even this very small amount of limited MWA data reaffirms the coronal diagnostic capability of low radio frequency emission and provides an early glimpse of the nature of radio observations that will become available as the next generation of low-frequency radio interferometers come online over the next few years. The expected performance increase going from 32 to 128 tiles and from  $\sim300$ m to $\sim3000$ m baselines is likely to be dramatic, during the period when the Sun is entering the peak of the solar cycle over the next few years.

In a very important milestone paper, \cite{Williams-etal.2012} used the MWA prototype to observe two 50$^{\circ}$ diameter fields in the southern sky in the 110 MHz to 200 MHz band, in order to evaluate the performance of the prototype, to develop techniques for epoch of reionisation experiments, and to make measurements of astronomical foregrounds in this field. \cite{Williams-etal.2012} developed a CASA-based calibration and imaging pipeline for the prototype data, and used it to produce $\sim$15' angular resolution images of the two fields, performing a blind source extraction using the resulting confusion-limited images to detect 655 sources at high significance and an additional 871 lower significance source candidates. A comparison of these sources with existing low-frequency radio surveys was used to assess the prototype array system performance, wide field analysis algorithms, and catalog quality. The source catalog was found to agree well with existing low-frequency surveys in these regions of the sky and with statistical distributions of point sources derived from Northern Hemisphere surveys; it represents one of the deepest surveys to date of this sky field in the 110 MHz to 200 MHz band. 

In addition to the extraction of the science results from prototype array data, considerable work has been put into understanding the science performance of the final instrument.  For example, \cite{bea12} examines the planned tile locations of the final MWA and accurately calculates the array sensitivity to the Epoch of Reionisation power spectrum of redshifted 21 cm emission. The calculation takes into account synthesis rotation, chromatic and asymmetrical baseline effects, and excludes modes that will be contaminated by foreground subtraction. \cite{bea12} conclude that with one full season of observation on two fields (900 and 700 hours), the MWA will be capable of a 14$\sigma$ detection of the power spectrum along with slope constraints. 

\section{Description of construction progress for final MWA}

In early 2012, the 32 tile prototype array was de-commissioned in order to make way for the construction of the final 128 tile system.  Construction of the final system commenced in February 2012, initially with civil works to establish the infrastructure required to support the instrument.  Primarily this work consisted of trenching around the site to reticulate power and optical fibre.  Seven trenches radiating from a central ``hub'' have been completed, and power and fibre laid to 16 locations on these trenches.  The 16 locations are the sites where receivers will be placed on concrete pads.  Capacity exists to support 32 receivers, although we will only initially utilise 16 receivers for 128 tiles.  Infrastructure capacity for expansion therefore exists, or else could be used to support other instrumentation.  This infrastructure work was completed in June 2012 and all electrical and data connections have been completed between the field hardware and the CSIRO central processing facility.

In parallel with the infrastructure work, accurate positions for the 128 tiles were surveyed and marked in May 2012.  Ground screens for all 128 tiles were constructed and placed at these locations in May 2012 and the 4,096 dipole antennas were constructed and installed on the ground screens in June 2012.

Receiver production at Poseidon Scientific Instruments (PSI) commenced in early 2012, with the first receivers delivered in March 2012.  These first receivers went through extensive EMC testing, to ensure that their performance complies with the strict radio quiet regulations for equipment fielded at the MRO.  It is expected that receivers will be deployed to the MRO starting in July 2012, at a pace determined by the production process at PSI.

In late 2011, the MWA project took delivery of a high performance computing cluster of 24 IBM iDataPlex servers, each housing two Nvidia GPGPUs.  This cluster, along with the already procured dedicated hardware and a data capture system known as 2PiP developed by the Smithsonian Astrophysical Observatory, forms the platform for the MWA correlator.  The correlator GPU code, the ``Harvard X-engine" is described by \cite{Clarketal}.  The IBM hardware, the FPGA hardware and firmware, and the GPU code have been integrated and tested in the laboratory environment and performance has been verified prior to installation at the MRO.

The installation of the correlator and other MWA signal processing hardware in the CSIRO central processing facility is expected to occur in the coming months, as the facility is handed over from the MRO construction contractors to CSIRO.  After this point in time, all the MWA signal processing hardware will be installed and integrated in the field.  Before our hardware can be integrated in the CSIRO building, we have the capacity for a temporary installation to commence commissioning using the infrastructure previously use for the 32 time prototype.

Thus, with infrastructure, tiles, receivers and correlator progressively installed in the field over the next few months, engineering commissioning of the full signal path for a fraction of the system (32 tiles at a time) can commence from approximately August 2012, with science commissioning of these 32 tile partial arrays commencing following engineering handover.  As receivers are progressively fielded and more tiles have their signal paths completed, engineering and science commissioning will proceed with 32 tile segments of the array, with practical completion (full instrument installed and commissioned as 4 x 32 tiles sub-arrays) reached in November 2012.  Following practical completion, final commissioning of the full instrument as a single 128 tile array can commence in early 2013.

Data from the MWA will be transported to the Pawsey HPC Centre for SKA Science in Perth via a dedicated 10 Gbps network.  This network is expected to be active from near the end of 2012.  Initial commissioning data may be transported between the MRO and Perth on disk, until the network link becomes active at sufficient capacity to support large-scale transfers.  Data will be stored in an archive in Perth, initially during commissioning on a 500 TB storage facility based at iVEC.  Later in 2013, as the Pawsey Centre is fully commissioned, the MWA will have access to a data store that will grow to 15 PB over a five year period (corresponding to approximately 300 days of MWA observing at maximum data production rate).  The data store will include real-time images as well as visibility data for reprocessing in the future.

Finally, software development for the project is coming to completion, with the Monitor and Control system in the final stages of development and testing, and the Real-Time System (RTS) fully tested and ready for integration with the system in an operational mode.  The RTS will utilise excess compute capacity on the IBM iDataPlex cluster that implements part of the correlator.

\section{Discussion}

With good progress in the final construction phase of the MWA and the valuable experience of the 32 tile prototype instrument behind us, we are well positioned to take advantage of an excellent facility, located in a prime region for low frequency radio astronomy.  As the MWA project moves into an initial commissioning phase, with a science commissioning team of 19 individuals from 10 institutions over 4 countries, it is likely that good science data will be obtained even before the point of practical completion is reached at the end of 2012.

With the SKA site decision now behind the community, assigning the Phase 1 and Phase 2 SKA low frequency arrays to the Australian site, the MWA can potentially become a very important facility during the SKA pre-construction phase.  As the only large low frequency interferometer at the MRO, the experience gained by undertaking challenging pathfinder observations in this frequency range will inform SKA low frequency design and specification activities.  Valuable experience will be gained in an operational sense, in the practical matters of maintaining a low frequency array in the environment identified with future SKA low frequency instrumentation.  Finally, and perhaps most importantly, the final MWA has available twice as much power and data infrastructure than is required to support the 128 tile array.  The MWA can provide this excess infrastructure to the SKA low consortium, consisting of: power and data connections existing to 16 locations in the field; connection to the CSIRO 160 dB shielded room, with approximately 10 spare water cooled racks to accommodate signal processing equipment; capacity on a 10 Gbps data link from the MRO to iVEC/Pawsey in Perth.  Thus, for prototyping small arrays of SKA low frequency antennas, the MWA can offer a ready-to-use signal path from the field at the MRO, to data storage in Perth.  Effective use of the excess MWA infrastructure could allow a very rapid and flexible deployment of SKA low prototype instrumentation at the MRO, in the environment that will eventually host SKA Phase 1 and Phase 2 low frequency arrays.  Furthermore, it would be possible to correlate the SKA low prototype antennas against the well-tested and operational MWA tiles.  Such a situation would likely allow an improved calibration of both the SKA antennas and the MWA tiles, to the mutual benefit of both projects.

Thus, the connection between the MWA and SKA low, if used to best effect, could be a substantial advantage in realising what is an ambitious timeline for the SKA pre-construction phase and Phase 1 construction.

\section*{Acknowledgments}
We acknowledge the Wajarri Yamatji people as the traditional owners of the Observatory site.  Support for the MWA comes from the partner institutions and funding agencies in each of the four MWA consortium countries.

\end{document}